\documentclass[preprint,authoryear,12pt]{elsarticle}
\usepackage[]{natbib}
\usepackage{amssymb}
\usepackage{fancyhdr}
\usepackage{courier}
\usepackage[normalem]{ulem}
\usepackage{float}
\usepackage{geometry}
\usepackage{epsfig}
\usepackage{color}
\usepackage{makeidx}
\usepackage{eurosym}
\usepackage{rotating}
\usepackage[stable]{footmisc}
\usepackage{multicol}
\usepackage{epsfig}

\def\mbh{$M_{\rm BH}$\/}
\def\nh{$n_{\mathrm{H}}$\/}
\def\lledd{$L/L_{\rm Edd}$}

\def\rfe{$R_{\rm FeII}$}

\def\feiiq{{\rm Fe}{\sc ii}$\lambda$4570\/}

\def\ltsima{$\; \buildrel < \over \sim \;$}
\def\ltsim{\lower.5ex\hbox{\ltsima}}  
\def\gtsima{$\; \buildrel > \over \sim \;$}

\def\gtsim{\lower.5ex\hbox{\gtsima}}

\def\civ{{\sc{Civ}}$\lambda$1549\/}
\def\civnc{{\sc{Civ}}$\lambda$1549$_{\rm NC}$\/}

\def\cm3{cm$^{-3}$\/}
\def\hb{{\sc{H}}$\beta$\/}
\def\hg{{\sc{H}}$\gamma$\/}
\def\hbbc{{\sc{H}}$\beta_{\rm BC}$\/}

\def\oiiiopt{{\sc{[Oiii]}}\-$\lambda\lambda$\-4959,\-5007\/}
\def\o4363{{\sc{[Oiii]}}$\lambda$4363\/}

\def\caii{{Ca{\sc ii}}}

\def\feiiuv{{{\sc{Feii}}}$_{\rm UV}$\/}
\def\feiiopt{{Fe \sc{ii}}$_{\rm opt}$\/}
\def\feii{{Fe\sc{ii}}\/}

\def\fe{{\sc{Fe}}\/}
\def\gs{{$\Gamma_{\mathrm{soft}}$\/}}

\def\fe76087{{\sc [Fe vii]}$\lambda$6087\/}
\def\oiii{{\sc [Oiii]}$\lambda$5007}

\def\kms{km~s$^{-1}$}

\def\rk{$R_{\rm K}$\/}
\def\ergss{ergs s$^{-1}$\/}

\def\rk{{$R{\rm _K}$}\/}
\def\heii{{{\sc H}e{\sc ii}}$\lambda$4686\/}

\def\apj{ApJ}
\def\apjl{ApJL}
\def\apjs{ApJS}
\def\pasj{PASJ}
\def\aj{AJ}

\def\mnras{MNRAS}
\def\aap{AAp}
\def\nat{Nat}

\usepackage{tocloft}
\defcitealias{shenho14}{SH14}
\begin{document}

\begin{frontmatter}
\title{\bf\sc Quasars  in the 4D Eigenvector 1 Context: \\A stroll down memory lane}
\author{Jack W. Sulentic\footnote{Instituto de  Astrof\'{\i}sica  de Andaluc\'{\i}a (CSIC), Granada, Spain.},    Paola Marziani\footnote{INAF, Osservatorio Astronomico di Padova, Italia.}}
\address{}
\date{}

\begin{abstract}
Recently some pessimism has been expressed about our lack of progress in understanding quasars over 
the 50+ year since their discovery. It is worthwhile to look back at some of the progress that has 
been made -- but still lies under the  radar -- perhaps because few people are working on optical/UV spectroscopy in this field. Great advances in understanding quasar  phenomenology have emerged using eigenvector techniques. The 4D eigenvector 1 context provides a  surrogate H-R Diagram for quasars with a source main  sequence driven by Eddington ratio convolved with line-of-sight orientation. Appreciating the striking  differences between quasars at opposite ends of the main  sequence (so-called population A and B sources) opens the door towards a unified model of quasar physics, geometry and kinematics.  We present a review of some of the progress that has been made  over the past 15 years, and point out unsolved issues. 
\end{abstract}
\end{frontmatter}
 
\section{Introduction}

The first fifty years of research on quasars has lead us to a paradigm involving 
three main (unresolved) components 1) supermassive black hole (SMBH; \citealt{zeldovichnovikov65,salpeter64}, 2) an emitting accretion disk (AD; \citealt{shields78,malkansargent82}) and 3) an obscuring torus \citep{antonuccimiller85}.  Component 3 arises from   attempts to unify Type 1 and 2 sources on the basis of orientation. A  4th component  would involve narrow line emission and sometimes radio-jets extending on scales of parsecs,  kpc or even Mpc along the rotation axis of the disk, resolved or partially resolved in the nearest sources   \citep{capettietal96,falckeetal98}. The AD may contribute to broad line emission and as such may be a constituent of the  broad line region (BLR) \citep{chenetal89,dumontcollinsouffrin90d,eracleoushalpern03,bonetal07},  while Component 4 involves only narrow-line region (NLR) emission unless the source is radio-loud. Cartoons showing the different components neatly aligned
are likely too optimistic but we can hope it is often the case. Moving beyond this 
paradigm has not been easy leading to some expressions of frustration.  Impediments to 
real progress involve lack of a clear definition of a quasar and the lack of any paradigm 
in which to contextualize source commonalities and differences. For a long time then 
we have been stuck with the paradigm as definition and the assumption that all quasars (or, 
better said, active galactic nuclei AGN) are the same. We have been waiting for one of 
the self-styled `elite" to lead us out of the darkness.

During the past 15 years, impatient with waiting,  we have assembled a formalism designed to 
contextualize  quasar diversity and identify the principal physical drivers of that diversity. 
This 4D Eigenvector 1 (4DE1) parameter space represents a surrogate H-R Diagram for type 1 
AGN. We extend the definition of Type 1 to include sources that show both broad-line  emission
from the principal optical and UV permitted lines and sources that show optical FeII emission 
and that are therefore expected to accrete at a moderate to high-rate (dimensionless accretion 
rates $\gtrsim 10^{-3} - 10^{-2}$, \citealt{woourry02,marzianietal03b}).  4DE1 was built upon 
pioneering studies of optical \citep{borosongreen92}, UV \citep{gaskell82} and 
X-ray \citep{wangetal96} spectra. By the year 2000 enough data and ideas were in place to introduce 
the 4DE1 formalism and the idea of two quasar populations A \& B \citep{sulenticetal00a,sulenticetal00b}
to emphasize the many differences between high and low accreting Type 1 sources.
The populations may represent two distinct quasar classes -- only one capable of 
radio-loudness \citep{zamfiretal08}--or opposite extremes  of a single quasar ``main sequence".

Developments in the statistical analysis before late 1999 are reviewed in \citet{sulenticetal00a}. 
Here we retrace developments that came mainly afterwards, since the turn of the century, and 
that are associated with a better understanding and expansion of the original eigenvector 1 results. 
A recent work by \citet{shenho14}  (hereafter \citetalias{shenho14}) basically confirms many past 
results and hopefully reawakens interest in the subject. The aim  of this paper is review the main 
steps in the development and exploitation of 4DE1 mainly by the group of Sulentic and collaborators 
(\S \ref{2001} -- \ref{2004}). We also review the ``rediscovery" paper \citetalias{shenho14} 
(\S \ref{2014}) and mention its followup \citep{sunshen15}. We maintain a loose chronological order 
(\S \ref{2001} -- \ref{2004}) in an attempt to clarify which results can be considered established  
and which still lie on uncertain ground-- therefore necessitating further study. We 
finally point out some major open issues 
(\S \ref{open}). 

\section{2000: Formulation of an Eigenvector 1-based parameter space}
\label{formulation}

The  4DE1 parameters introduced in 2000 involve: 1) full width half maximum (FWHM) of broad  \hb\ 
(\hbbc); 2)  equivalent width (EW) ratio of the optical \feiiq\ blue blend and broad \hb\ (\rfe). 
The choice of equivalent width was motivated by its widespread availability in low $z$\ ($<$0.7) spectra.
In more recent time we have considered \rfe\ as defined from the intensity or flux ratio in order 
to avoid division by a continuum  that is often steeply rising toward the blue; 3) profile shift at 
half maximum  of high ionization \civ\, c(1/2) and 4) soft X-ray photon index (\gs). The parameters
are thought to be: 1) a measure of virialized motions in a low-ionization line emitting AD 
or flattened system of clouds that is considered an important virial estimator of black hole mass for 
large samples of quasars. 2) A sensitive diagnostic of  ionization,  and column density in  BLR  gas 
arising, as far as we can tell from shielded or outer parts of the BLR. The strength of optical FeII 
emission in many sources has been argued to support the AD origin for the emission \citep{perryetal88,dultzinhacyanetal99,jolyetal08} and a role for metallicity. 3) A strong diagnostic of winds/outflows in the  higher ionization broad line gas. 4) a  diagnostic measure of thermal emission likely connected with the accretion disk, and to the accretion state \citep[e.g.,][although see \citealt{doneetal12} for a dissenting view]{mineshigeetal00}. These measures were chosen because they were available for: 1) large numbers (100+) of low $z$\ quasars with 2) high S/N spectra 
(S/N$>$20) and 3) because they showed statistically significant dispersion along the 4DE1 main sequence.  They are Eigenvector 1 because they are strongly correlated and if parameters 3 and 4 had been included in the first PCA analysis they would have contributed much of the power of Eigenvector 1. They  are 
``orthogonal" in the sense that  they involve parameters describing independent aspects of quasar  phenomenology as well as different physical processes connected to the BLR.

One can mark 1998 as the year when theoretical attempts to model the BLR (of well studied NGC5548) were acknowledged to have failed at least in part \citep{dumontetal98}. We think the pessimism  that followed is reflected in the incompatibility of measures for population A and B quasars (e.g.  significantly different ionization parameter $U$\ and geometric factor $f$). After 1990, there has been a widening void between theory and observations, probably because the large diversity of quasar properties has not been taken into account as needed.   Much more success is  likely to arise from attempts at modelling Pop. A (e.g. I Zw 1) and B (NGC 5548) or individual spectral types (as defined by \citealt{sulenticetal02}; \S \ref{2002} \ below) separately.  Then the question of uniting the two solutions -- or the need to invoke separate  quasar populations (perhaps driven by a critical accretion rate, \citealt{marzianietal14}) can be addressed.

\section{2001: The Physical Drivers of 4DE1}
\label{2001}
\begin{figure}[htp!]
\centering
\includegraphics[width=6.1in]{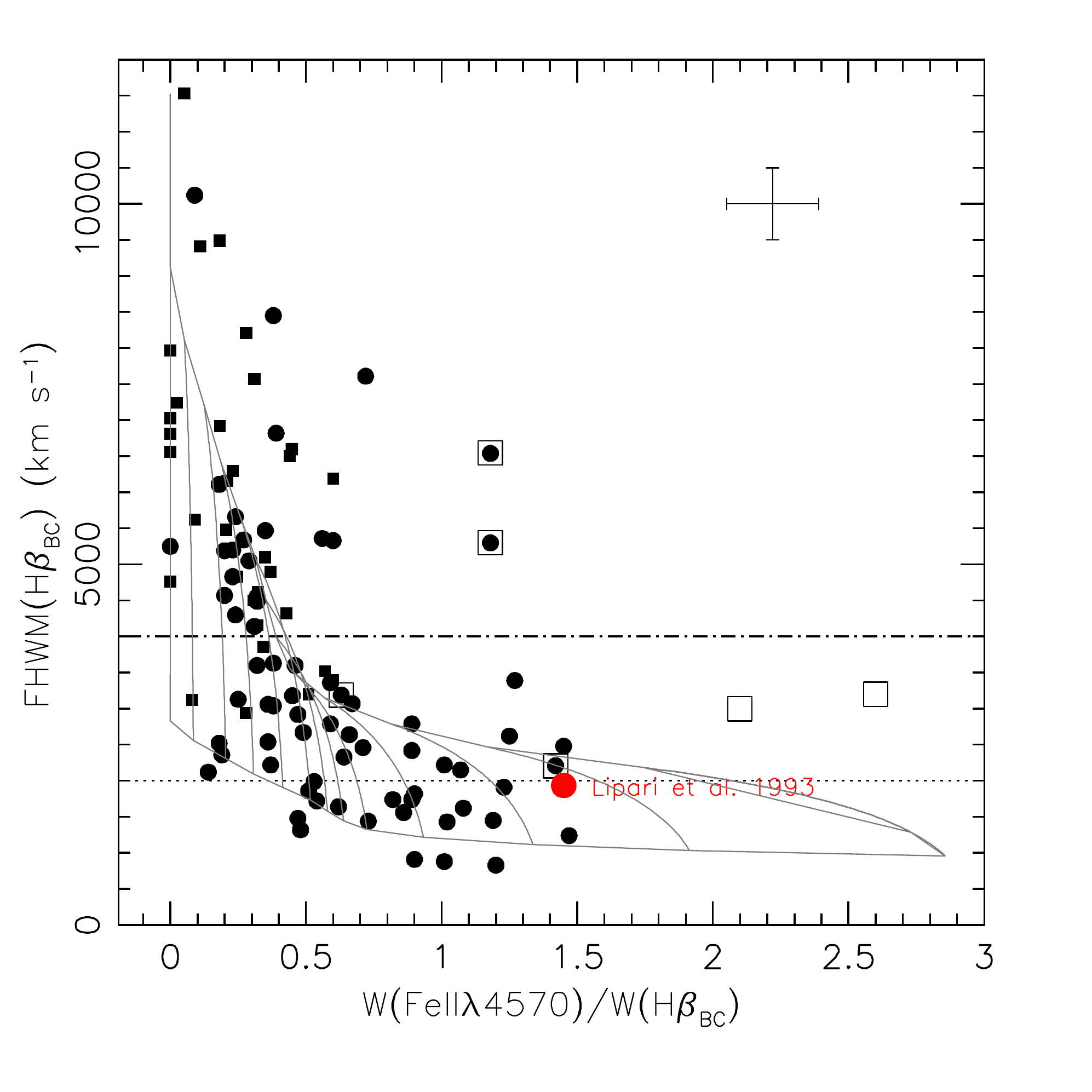}
\caption{Simplified version of  Fig. 4 in  \citet{marzianietal01}. The 4DE1 optical plane shows sources from  the \citet{sulenticetal00b} sample with a superposed grid of 
theoretical values of viewing angle $i$ (10 $\le i \le $ 40) and $L/M$, expressed in solar 
values, for 3.1 $\le \log M \le 4.5$ (implying $\log$  \lledd $\approx$ 0), at steps 
of $\delta \log L/M$ = 0.1.  A value of $\log M \approx 8$\ in solar units was assumed 
for $U$.   Filled circles and squares represent  RQ  and RL sources respectively in the \citet{sulenticetal00b} sample. Boxes identify BAL QSOs (2 of them, Mark 231 and IRAS 0759+651 not in  \citealt{sulenticetal00b}).   The red spot mark the average for the strong \feii\ emitters of \citet{liparietal93}. Error bars  in the upper right corner indicate typical 2$\sigma$\ uncertainties for a data point at  FWHM  $\approx$ 4000 \kms\ and \rfe $\approx$ 0.5.\label{fig:m01}} 
\end{figure}

After we proposed the 4DE1 contextualization we began to search for the physical drivers. We knew
that, unlike stars, quasars are very unlikely to show the same spectroscopic properties at different 
viewing angles \citep{willsbrowne86}. \citet{borosongreen92} suggested \lledd\ as a possible physical 
driver by exclusion, as a ``best guess". Not source luminosity (an Eigenvector 2 parameter in their PCA), 
not simply orientation (the extreme range in FWHM alone H$\beta$ precludes that) nor geometry
--all were discussed much more. They were writing in the early 1990s using a sample of 87 quasars 
so their discussions are perhaps out of date from todayís perspective. More then two measures
would likely be needed to break the degeneracy between source orientation and physics. The following year \citep{marzianietal01} we explored the physical drivers of source occupation  along the optical ``main sequence" and concluded that source orientation ($\theta$) convolved with the ratio of quasar luminosity 
to black hole mass ($L/$\mbh $\propto$\ Eddington ratio, as anticipated by \citealt{borosongreen92}) 
could describe rather well source occupation in the optical plane of 4DE1. (Fig. \ref{fig:m01}) reproduces
such an overlay on our heterogeneous but high S/N sample (see also ESO Messenger for June 2001  \citealt{sulenticetal01}).  {{It was rejected by the Nature editor.}}  The following year two of us considered the role of black hole mass over $\approx$ 8dex  range \citep{zamanovmarziani02}. This was accompanied by an attempt
to estimate \lledd\ from $L$ and \mbh, the latter estimate from line width and scaling relations 
\citep{laor00,boroson02,marzianietal03b}.  The US author of this paper explained to his collaborators (from, Italy,  Mexico and Slovenia) that these results would remain largely unacknowledged until someone from an ``elite"  institution rediscovered them. That's how it works in the US at least. Many times over the past 14 years  he  lamented the slowness of his elite colleagues to grasp the significance of these really quite simple results. Finally it has come to pass \citepalias{shenho14}! And with the expected endorsements and encomia.

At this point it might be useful to compare the abstract  of  \citetalias{shenho14} with that of  \citet{marzianietal01}. In one sense, \citetalias{shenho14} is ``one step forward and one step back".  The new paper uses a much larger  sample of quasars drawn from SDSS (one step forward) but, most of the spectra that they use  and measure \citep{shenetal11} are unsuitable to produce an equal or greater description of the role
for L/L$_{Edd}$  (one step back). SDSS is a gold mine but at the same time it is a minefield for the uninitiated user. Automated processing of these spectra is indeed a walk through a minefield. We have emphasized since 1996 the need for, and importance of, quasar spectra with high enough S/N and resolution to permit reliable spectroscopic parameter measures. That is why \citet{borosongreen92} was such a step forward
and their measures are still used 25 years later.

Although in some places it may be treasonous to point it out, most low S/N quasar spectra are not 
suitable for useful spectroscopic measures.  If this statement were untrue then, in the context of future quasar studies, most of the justification for 10m class telescopes -- beyond observing, sure to be proposed,  {obscured}  quasars --  would be  negated. Figure \ref{fig:shz} shows a comparison of automated 
measures of SDSS quasars \citep{shenetal11} used in the SH14 study  with IRAF SPECFIT  measures of \citet{zamfiretal10}. The comparison involves the brightest quasars in SDSS-DR5 and therefore generally 
showing S/N$>$20. Continuum S/N not to be confused with S/N computed in regions with strong emission lines. Surprisingly many of the FWHM H$\beta$\ measures used in this paper \citep{shenetal11} are also strongly discordant with more  detailed  analysis presented a few years ago \citep{zamfiretal10} for the 500 brightest DR5 quasars (Fig. \ref{fig:shz}). This is especially true for so-called population B quasars with FWHM H$\beta > $4000 \kms\ as was shown in 2002  \citep{sulenticetal02}. Figure 3 in \citet{marzianietal03a} quantifies the dependence of \feii\ detectivity on S/N and FWHM H$\beta$.  Each Shen measure is connected to the corresponding Zamfir measure. Taken at face value, automated measures can even mislead statistical analyses i.e., it is not that they fail occasionally because of the usual problems but they create categories of spurious {\em classes of sources} especially for large sample sizes.  {This is shown by the left panel of Fig. \ref{fig:shz} where the ``outliers'' in the optical plane of 4DE1 turn out to agree with the main sequence once more accurate measures are used.}  An additional problem  is  that \heii\ emission is apparently included as part of the  \feii\ blue blend in the \citet{shenetal11} data. \heii\ measurements are not reported there.  \heii\ is pernicious leading to overestimation of \heii\ strength, overestimation of  \feii\ template width and claims of an \feii\ redshift relative to H$\beta$\ \citep{huetal08,sulenticetal12}.  {This is shown by   Fig. \ref{fig:heii} where the one \feii\ strong source is more straightforwardly interpreted in terms of weak \feii, and strong and very broad \heii. }

The new analysis might not have been ``two steps back" if all quasar spectra were the same. Our work from 1989-2001 also emphasized the diversity of the spectroscopic (and other)  properties of quasars. The concept of two quasar populations (A and B) was introduced  to further emphasize this point -- it is a simplified   analogue of the seven principal spectral types identified in the stellar H-R Diagram. Naturally  low S/N spectra tend to diminish this diversity leading  to the quite natural temptation to indiscriminately average many noisy spectra together in order to produce higher S/N composites. But without a context with which to average,  these composites it confuses rather then clarifies the quasar phenomenology and raises an impediment to improved physical models. What would a  composite of OBAFGKM stellar spectra give us?

Figure 1 of \citetalias{shenho14} allows us  to visualize the main E1 trend but it does not improve our understanding of  the distribution of sources in the 4DE1  optical plane because the \feii\ blends, 
present in virtually all quasar spectra, {cannot be properly  detected, much less modeled, in 90-95\%\ of 
the SDSS spectra}. A continuum is likely fit on top of  noisy \feii\ emission resulting in a zero, or too low, measure of \feii\ strength. A censored data analysis is appropriate in this context \citep{sulenticetal02}. 

\section{2002 -- Eigenvector Binning and First Explorations of \oiii}
\label {2002}

\subsection{Defining spectral types}

We are still speaking about pre-SDSS years. Further exploration/exploitation of 4DE1 required a  larger sample but fewer then 200 quasars with moderate/high S/N spectra existed at this early time.  While engaged in obtaining new data with 2m class telescopes in Spain, Mexico, Italy and Chile we realized that another approach (following OBAFGKM) would be to bin the optical plane in order to better  contrast quasars along the 4DE1 main sequence. Bin sizes were chosen so that all quasars within a given bin were statistically indistinguishable. If we were to bin stellar spectra we would bin by spectral type or even subtype since indiscriminate averaging will obscure any physical insights. Most of our papers showed that tremendous spectral diversity existed in the type-1 quasar population. We chose bins of equal size because we have no evidence that the physical drivers of 4DE1 source occupation correlate with measured properties in a way to justify unequal bins. We are not yet in a position to know for sure. \citetalias{shenho14} rediscover binning but with bins of unequal size which seems to us a step back. Binning should be predefined rather than driven by a distribution in a parameter space defined using low S/N spectra. Our first paper showing average spectra  of H$\beta$\  \citep{sulenticetal02} revealed a clear change in the profile shape along the sequence with profiles for sources with lowest inferred Eddington ratio showing an extra very broad and redshifted component (VBC). This is not to say that the binning  done by \citet{sulenticetal02} isolated sources that were scattering randomly around an average value in each bin: rather, the grid of \citet{marzianietal01} clearly shows that within each bin orientation and Eddington ratio trends were expected.

\subsection{Very broad component and very broad line region}

This and other striking differences motivated us however to separate quasars into two populations: population A, involving  mostly radio-quiet (RQ) high accretors with symmetric Lorentz-like H$\beta$ profiles, strong optical \feii\  emission, a \civ\ blueshift / asymmetry and a soft X-ray excess. Population B sources are lower accretors including most of the classical radio-louds quasars (RL). They show composite double-Gaussian (BC+VBC) H$\beta$\ profiles, weaker optical \feii\ emission, unshifted (or redshifted?) \civ\ and no soft X-ray excess. Some quasars show an unusually strong VBC component that dominates the H$\beta$\ profile e.g. \citep[PG 1416--129][]{sulenticetal00c}.  At first the BC + VBC decomposition was purely phenomenological: but
extreme VBC sources give us a clearer insight into the shape of \hb\ VBC.  In most cases it was possible to reproduce H$\beta$ well for the overwhelming majority of population B sources and specifically the ones classified as AR,R or AR,B (i.e., redward asymmetric, AR, with red or blueshifted peak, \citealt{sulentic89}). A very broad line region (VBLR) has been postulated since the mid 1980s and was initially suggested to be due to optically thin gas in the innermost BLR \citep{petersonferland86,zheng92,shieldsetal95,morrisward89}.  Optical ``thinness" has strong implications for the maximum luminosity associated with a line: if the medium is optically thin the intensity of the same recombination line is governed by the volume and density of the emitting gas and is not directly related to the luminosity of the ionizing continuum \citep{marzianietal06}. 

 {Earlier observational definitions were based  mostly on the \civ\ profile. \citet{brothertonetal94a} postulated the existence of a intermediate line region emitting most of \civ\ of FWHM $\approx$  2000 \kms, and a broader component (VBC) to account for the extended \civ\ wings. \citet{corbin95} interpreted the redshift often observed in the \hb\ wings as a gravitational redshift affect the VBC. By 2002-2003 it was clear that the profile analysis  of the \civ\ profile was more ambiguous  than the one of \hb. While \hb\ customarily shows a narrower component sharply separated from the broad component, the \civ\ profile of Pop. B sources shows a prominent semi-broad core that merges smoothly with the line base. We already had suggested an interpretation in terms of density stratified NLR  to account for the \civ\ core profile \citep{marzianietal96,sulenticmarziani99}.  }

 {  No VBC is apparently present in \feii: we run several tests and we concluded that the VBC cannot be as strong in \hb\ and, if present, is too weak to be appreciable. The BC and VBC decompositions therefore reflects the ionization stratification within the BLR, simplifying what is probably a continuous trend. Two main regions are identified: a higher ionization VBLR, associated with the VBC (i.e., line base and line wings), and a lower one associated with the BC, that emits most of all of \feii. The absence of \feii\ emission can be considered  a defining property of the VBLR. }

 {  It was later shown that the VBC is strong in luminous quasars, and that its luminosity cannot be explained by optically thin gas. Presently, the VBLR is understood as inner emitting region whose high ionization degree leads to a lower responsivity to continuum changes \citep{koristagoad04,goadkorista14}.}




\subsection{Blue outliers: linking wind signatures on widely different  spatial scales?}

During this year we also began exploration of \oiii\ as a diagnostic of the narrow line region. This kind of study requires very high resolution. In fact BG92 expressed frustration with attempts to parameterize \oiii\  using standard measures like equivalent width (EW). The reason for this was that even the very good spectra employed in BG92 were too low resolution to allow realization that there are two distinct components of \oiiiopt.  We obtained spectra at San Pedro Martir, Calar Alto and KPNO and identified a class of ``blue outlier''  sources \citep{zamanovetal02} where the velocity separation between the two \oiii\ components (unshifted+ blueshifted) was largest, and the profile appeared with a blueshift $> 250$ \kms\ in amplitude. They appear to favor population A   quasars  (Fig. \ref{fig:bo})which also show large \civ\ blueshifts, suggesting a kinematic linkage between the narrow and broad line regions.  Further examples were presented the next year \citep{marzianietal03b}, and the existence of large blueshifts and semi-broad \oiiiopt\ emission has been confirmed in a number of studies  \citep{aokietal05,bianetal05,komossaetal08,zhangetal11}. Recent studies even find  emission extended on galactic scales giving rise to an integrated semibroad profile \citep{canodiazetal12}, although the original results of \citet{zamanovetal02} indicated that the semibroad \oiiiopt\ component originated in a very compact NLR, $\sim$ 1 pc in size. At that time and with our data the question mark on the BLR-NLR linkage was wise.  Nowadays it seems that quasar outflows are linked from the BLR up to the circumnuclear regions, where molecular outflows are detected, at least in the extreme source Mrk 231 \citep{feruglioetal15,tombesietal15}, a BAL QSOs that was noted in our early analysis because of the abnormally large \hb\ FWHM that placed the source outside the main sequence of 4DE1 \citep{marzianietal01,sulenticetal06}.   It is unfortunate that the binning in SH14 was nonuniform because it damped out some of the \oiii\ profile diversity in their Figure 2. Bin size should not be motivated to maximize the number of quasars within a bin but rather to explore the diversity of Type 1 spectroscopic properties within the parameter space.  {We have evidence that even our already rather fine  subdivision should be made even finer.} Given the complexity of \feiiopt\ emission and weakness of \oiiiopt\ in extreme population A sources one must be especially cautious interpreting results for these quasars. 

The essential point  that emerged in 2002/2003 is that \oiiiopt\ shows two components. The narrower unshifted component shows a strong change along the 4DE1 main sequence becoming very weak or absent in extreme Pop. A  (highest \lledd) sources. A second semibroad and blueshifted component may be  present in all sources but is most prominent in extreme population A where  it  shows blueshifts up to $\sim$ 1000 \kms\ (I Zw 1 is a famous example of the so-called blue outliers). This is not  a novelty of \citetalias{shenho14}, since this trends was discussed in \citet{marzianietal06} on the basis of the previous 2002-2003 papers.  The semibroad component is difficult to study  in many sources requiring spectra of very high S/N and resolution. In some of these rare spectra \oiiiopt\ is  fully resolved into two components \citep{grupeetal99}. A further problem mentioned in some of the above references involves slit effects connected with extended emission line region (EELR)   contaminating the spectrum.  


\section{2003: an useful, uniform dataset after \citet{borosongreen92}} 
\label{2003}

\subsection{A physical dichotomy between Pop. A and B?}

This was ``a very good year" when observations in Mexico, Spain, Italy and Chile (ESO) enabled us to expand  our sample to more than 200 low $z$\ quasars. This was a big deal on the eve of the SDSS. Even better we began to explore in more detail of role of black hole mass and Eddington ratio along the quasar  main sequence. We were not the first \citep{marzianietal03b} but the 4DE1 approach allowed us to  gain some insight, and to suggest that the separation between Pop. A and B occurred as a fairly well defined \lledd, $\approx 0.1 - 0.2$\ for a quasar with $\log$ \mbh$\sim$8. It is possible that this critical Eddington ratio may signal the transition to a slim disk from an optically thick, geometrically thin disk \citep{abramowiczetal88}.  


Are population A and B simply extreme ends of a main sequence or do they represent two distinct quasar populations? This remains an open question as far as we are concerned, although  there is now evidence that the A/B differences could be associated with an accretion mode transition, with higher \lledd\ sources accreting in an advection dominated  accretion flow (ADAF)  \citep[][and references therein]{marzianietal14}.   The gravitational explanation for the large redshift observed at the line base of \hb\ was found to be too large to be consistent with the assumption of predominantly virial motion \citep{marzianietal03b}.

\subsection{The RL/RQ dichotomy}

It was also the year when more detailed studies of the RQ -- RL dichotomy were explored in the 4DE1 context first  with our own atlas \citep{sulenticetal03} and later with the 500 brightest SDSS DR5 quasars \citep{zamfiretal08}.  The strategy here was to explore the distribution along the 4DE1 main sequence of the most unambiguous  class of RL quasars--those showing classical double-lobe (LD) radio morphology. They  clearly show a strong concentration at the low \lledd\ (Population B) end of the main  sequence. Almost all show a consistent absence of a soft X-ray excess and CIV blueshift.  The restricted 4DE1 parameter space occupation of LD RL sources is perhaps the strongest evidence for  a physical dichotomy between RL and RQ  quasars (Fig. \ref{fig:z08}) -- if 4DE1 parameters reflect  fundamental aspects of BLR structure and kinematics.  The situation is much less clear  for core-dominated RL -- a few are interpreted as preferentially aligned LD sources but the majority especially  those weaker than the LD sample lower limit ($\log P_\mathrm{1.415 GHz} \approx$ 31.6 \ergss Hz) have been attributed to diverse scenarios (e.g. ``young" \citealt{vanbreugeletal84} and ``frustrated" \citealt{fantietal95} scenarios). If these scenarios  reflect reality they imply that some CD sources above ($\log L_\mathrm{1.415 GHz} \approx$ 31.6 \ergss Hz$^{-1}$) may not be RL in the classical sense. They will rise out of the RQ  population and many will become rapidly quenching flat-spectrum  radio sources, or eventually cross the LD boundary   producing LD morphology    (as proposed for the  compact steep spectrum (CSS) radio sources).  It was shown in \citet{zamfiretal08} that CD sources in the range $\log L_\mathrm{1.415 GHz} \sim$ 30.0-31.6 \ergss Hz$^{-1}$ distribute like RQ quasars and not LD RL. Looking at Fig. 4 and 5 of \citet{zamfiretal08} makes obvious that 4DE1 has much to say about the RL-RQ dichotomy.

\section {2004(+2007): interpreting the rest-frame UV spectrum and the \civ\ profiles along 4DE1} 
\label{2004}


Up to this point we were not in a position to explore the other 4DE1 parameters in much detail. Myriads of \civ\ spectra existed by this time including the very good spectra from the Palomar surveys  \citep[e.g.,][]{bartheletal90}   but we had no way to reliably estimate quasar rest frames. Since we knew that \civ\ profile shifts showed large diversity along the 4DE1 main sequence we could not proceed. We could not use existing studies \citep{brothertonetal94} because  their methods of \civ\ profile decomposition were incompatible with ours and not physically justified \citep{sulenticmarziani99}. We argued that a narrow component of \civ\ must be cautiously subtracted and that it was stronger in population B quasars, with semi-broad profiles.  {Our approach was motivated by lack of a well defined  critical density associated with \civ\ (instead suppressing \oiiiopt\ at density \nh $\gtrsim 10^{6}$\ cm$^{-3}$. The assumption of a density radial trend    gave additional arguments that \civ\ narrow component (NC) could be broader than \oiiiopt\ by up to 3 times following the simple modelization of \citet{netzer90}.  \civ\ emission it favored at high ionization parameter, with a steep decrease toward lower ionization conditions. The emissivity of \hb\ has instead a much flatter dependence on ionization parameter, and its volume integrated profile is weighted toward the outer emitting regions. Hence, the \civ\ NC profile may merge smoothly with the BC, while the \hb\ NC profiles stands out as an easily separable feature that is discontinuous from the BC. Since \civnc\ is due mainly to the inner NLR, \civnc\ blueshift within a few hundreds \kms\ per second are expected   \citep{zamanovetal02,aokietal05,boroson05,bianetal05,komossaetal08}.  }

We also gave a recipe for estimating if NC is present and how to subtract it.  Obviously if one does not distinguish population A and B things look more confusing. If \civ\ NC is present and you do not subtract then you measure \civ\ too narrow, too strong and  probably less shifted. \citet{baskinlaor05b} made a  NC subtraction although it was much lower for  sources in common. In population B, where the NC merges smoothly with the BC, an NC definition  is often ambiguous. The most controversial situation indeed concerns population B sources. There the BC   merges smoothly with the NC profiles, making the separation of NC and BC operationally ambiguous.  However, not subtracting the NC   may introduce a significant bias lowering \mbh\ mass estimates, and making them fortuitously consistent with the \hb-derived ones \citep{sulenticetal07}. 

Finally the number of low-$z$ spectra with \civ\ spectra in the HST archive passed 100 enabling us to produce binned spectra  \citep{bachevetal04,sulenticetal07}. Binning (and rest frame estimation) came from previously obtained matching optical spectra of the H$\beta$ region. Now it was very clear that \civ\ blueshifts were a Population A phenomenon likely associated with a disk wind or outflows in these highest accretors (Fig. \ref {fig:s07}).

This year also saw the beginning of our study of high redshift quasars in the 4DE1 context.  This required 8m class telescope time  because we were forced  to follow \hb\ + \oiii\ into the infrared in order to both assign high $z$\ sources to the  correct bin and to obtain a reliable estimate of the rest frame. We used VLT-ISAAC to obtain  spectra of unprecedented S/N for 52 high $z$\ sources. We could, for the first time, observe changes in 4DE1 source occupation and we could explore UV surrogates for estimating  rest frame and black hole mass -- lessening the need for IR spectra of large samples. This and  much more await rediscovery and we certainly hope it will not take another 15 years.

Equivalent measures for items 1 and 2 (\S \ref{formulation}) have  been identified in the redshifted UV spectra of high-redshift quasars \citep{negreteetal14}. In the simplest consideration  of quasar diversity  involving Population A and B sources (described below), virtually all multiwavelength measures of quasars show differences \cite[see Table 5 of][]{sulenticetal07}.





\section{2009} 

 {Until the mid 2000s our main results were based on the study of low-$z$, predominantly low luminosity sources. In 2009 we presented the largest instalment of IR spectra of Hamburg-ESO intermediate redshift quasars covering the \hb\ spectra range \citep{marzianietal09}. In total, considering previous batches \citep{sulenticetal04,sulenticetal06}, we had  data available for 52 very luminous quasars. Comparison with previous observations at lower $L$\ allowed as to start the analysis along the $L$\ dependent ``eigenvector 2.'' Perhaps the most remarkable result was the detection of the systematic increase in minimum-FWHM as a function of luminosity, a trend that is expected if: (1) the \hb\ line profile is virially broadened; (2) \lledd=1 is a physical limit; and (3) the BLR radius scales as $L \propto r^{\alpha}$\ with luminosity, with $\alpha \approx 0.5 - 0.65$ \citep{bentzetal13,kaspietal05}.  We cannot find high luminosity narrow line sources if these assumptions are valid. \citet{negreteetal12} and \citet{marzianisulentic14} have shown the equivalence of a broader-lined source with the prototype I Zw 1. This implies that the limit at FWHM = 2000 \kms\ for defining Narrow Line Seyfert 1 sources makes sense only at low-luminosity. Relaxing  this limit would allow to appreciate that there are sources which are analogous to local \feii\ strong NLSy1s in terms of physical condition but simply with broader lines. A more physically oriented criterion could attempt to isolate sources at \lledd $\rightarrow \mathcal{O}(1)$, which means applying the empirical criterion \rfe$\gtrsim$1\ (i.e., to isolate the sources we called xA in \citealt{marzianisulentic14}).    A second intriguing result was the high frequency of blueshifted, low equivalent width \oiii\ profiles. Since not all HE sources were high \lledd, we suggested that this trend could be associated with a NLR evolutionary effect (see also \S \ref{2014}). }

\section{2014, or the year of the final rediscovery}
\label{2014}


\subsection{Results involving \oiiiopt}

SH14 establish in a firmer way the sequence of \oiiiopt\ profile behaviour, something also described much earlier \citep{zamanovetal02,marzianietal03e, marzianietal06,marzianisulentic12}. Figure \ref{fig:bo}  supports the trend concerning shifts of \oiii\ shown in their Extended Data Figure 2. Its significance was discussed in the above references and  also by \citet{bianetal05,huetal08,komossaetal08}. 

The panel SFI of \citetalias{shenho14} shows the \oiiiopt\ ``Baldwin effect," i.e., a systematic decrease of \oiiiopt\ EW with $L$.  It has already been stressed \citep{netzeretal04}, and is consistent with the VLT data of Hamburg ESO quasars  \citep{sulenticetal04,sulenticetal06,marzianietal09}. It is interesting that \citet{shenho14} find a different  trend for the \oiiiopt\ blue wings -- in the sense that   wings remains prominent also at large luminosity, and shows little or no Baldwin effect. These finding may support an evolutionary interpretation of E1 \citep{dultzin-hacyanetal07,marzianietal14}: at the extreme population A end, small \mbh\ black holes accreting at high rate, with evidence of circumnuclear star formation \citep[e.g.,][]{sanietal10}; at the end of population B, mostly very massive  quasars associated with low accretion rate and old structures like extended radio-lobes of Fanaroff-Riley II sources.  The evolutionary interpretation seems straightforward for RQ quasars. However, while the properties of extreme population A and B are obviously different, and old radio sources are preferentially found at the extreme of population B \citep{zamfiretal08},  it is still not obvious which sources population A or B sources may be considered the radio loud progenitors of FRII in the 4DE1 sequence. A tantalising possibility is that compact-steep sources (CSS) may evolve into FRII \citep{vanbreugeletal84,sulenticetal15}.

\subsection{Physical interpretation of the optical E1 diagrams}

SH14 adopt an approach that is at first as surprising as it is unconventional. They consider that quasars hosting massive black holes -- associated with larger hosts -- should form in  denser environments. They conclude that quasars with higher \rfe\ are on average less massive because  they form in less dense environments. The deduction is intriguing and, as presented, has a statistical  value: it supports the idea that there is also a trend of mass with  \rfe\ that may lead to a  sequence of $L$/\mbh\ if the $L$\ range is not large. This would account for a displacement in the horizontal  direction i.e., \rfe, but not in the direction of FWHM. They then conclude that the spread in FWHM is not due to  \mbh\ and may largely be due  to orientation. As a proof it is incomplete. Their approach also  follows the  hypothesis that it is not possible to have strong \feii\ in very massive sources. However, this is contradicted by our VLT--ISAAC  observations of 52 high luminosity ESO-Hamburg quasars where we often find  strong  \feii\ \citep{sulenticetal04,sulenticetal06,marzianietal09} that led to even suspect an ``anti-Baldwin effect''  for \feiiopt.  {In our opinion, a more convincing proof is offered by  at $\sigma_{\star}$\ systematically decreasing with increasing \feii\ at fixed quasar luminosity. If \mbh\ and host galaxy bulge mass are tightly correlated \citep{ferraresemerritt00,gebhardtetal00}, then \feii\ strength increases with the $L/M$\ ratio \citep{sunshen15}.}

In order to test the effect of orientation SH14 consider the ratio between black hole mass 
(derived from the \mbh\ -- $\sigma_\star$\ relation) and the virial product  $f$ = $G M_\mathrm{BH,\sigma} / r \mathrm{FWHM}^2$. This is basically the structure factor 
that should be independent of line width if not influenced  by viewing angle. A dependence on orientation is however  expected from the assumption of a flattened, axially symmetric broad line region. They indeed find a strong  dependence and conclude that orientation strongly affects FWHM \hb. The orientation effect has been extensively explored,  and a strong dependence confirmed, using radio-loud quasars where orientation can be inferred from  radio morphology \citep{willsbrowne86,sulenticetal03,jarvismclure06,zamfiretal08,runnoeetal12,runnoeetal13}. This rediscovery adds evidence to the 
hypothesis that orientation is indeed vertically displacing sources in the 4DE1 
optical plane \citep[as clearly also seen in][]{marzianietal01}. It leaves open the possibility that  $f$ is changing due to other effects (for example Eddington ratio, although it is reasonable to expect that the effect of \lledd\ is not as large as that due to orientation). 

The next conclusion in SH14 is quite surprising: that orientation is driving the change in Balmer line profiles  between Population B (broader)  and Population  A (narrower). A significant vertical displacement is expected from \mbh\ (since FWHM $\propto \sqrt{M_\mathrm{BH}}$). \citet{zamanovmarziani02},  in a paper that complemented  \citet{marzianietal01} showed the effect of changing \mbh\ on the optical plane.  We also considered the hypothesis that orientation was the dominating factor in 4DE1 and that it was  driving Pop. A and B differences. We concluded that this is very unlikely: there are differences not only in  line profiles but also in diagnostic line ratios that cannot be reconciled with an orientation effect, and demand a change in physical conditions. The grid  in \citet{marzianietal01} -- derived also considering trends in diagnostic ratios -- was computed for a fixed  black hole mass. It showed that the effect of orientation is most important for sources with weak \rfe. If more  luminous sources are included we find a vertical spread associated with increasing mass: i.e., large \rfe\ and  large FWHM. Considering the main effects, a 4DE1 contextualization must involve at least 3 dimensions in  order  to distinguish between the effects of orientation and mass. And SH14 still results do not provide an estimation of viewing angle that is valid for individual sources. 

Accepting at face values the results of \citet{marzianietal01} and \citetalias{shenho14}, the conclusion would be that spectral type A1 (\rfe $\le 0.5$, FWHM $\le 4000$ \kms) is mainly due to face-on population B sources. Indeed an inspection of the sources in spectral bin A1 for the sample of \citet{marzianietal03b} reveal a composite distribution. Bin A1 seems to be populated by sources which could be population B oriented pole on (i.e., powerful CD radio-loud sources showing a relatively narrow BC and a faint VBC) but also that are the apparent extension of A2 with lower \rfe. The latter type slightly outnumbered the type B sources in our samples \citep[including the one of][]{zamfiretal10}, yielding a median A1 profile consistent with a Lorentzian function. 

\subsection{4DE1 parameters}



Lowest EW \civ\ sources colored black/blue are favoring the lower end of the main sequence in the  \rfe\ vs FWHM \hb\ plane. This rediscovered trend  in SH14 SFIV was previously reported in  \citet{bachevetal04} and \citet{sulenticetal07}. Note that sources in Population bin A1 do show relatively large W(\civ). \civ\ equivalent width is especially useful for identifying spectral bin type of high redshift quasars  \citet{negreteetal14a} discuss UV identification criteria in the 4DE1 context showing that
W(\civ)$\ltsim$ 50 \AA\ is a sufficient criterion for identifying extreme Pop. A sources (bin types
A2, A3 and A4 -- but not A1). \civ\ as well as intermediate ionization line widths and profile shapes 
distinguish bin A1 from B1.

Early work   resulted in the inclusion of \gs\ as a principal 4DE1 parameter. The presence of  a hard X-ray power-law may represent the best ``operational definition" unifying all AGN but  it was found to show less intrinsic dispersion than the soft X-ray photon index \citep{brandtetal97}.  There is a clear trend along the main sequence but SH14 can only rediscover in their SFVII what we have already learned \citep{wangetal96,sulenticetal00b}. It is the availability of X-ray spectra that limits 
our statistical analyses. The main sequence trend is clearest when comparing mean/median measures for  the upper (population A) and lower (population B)  ends. Spectral types A2-A4 show the largest  soft X-ray excesses and Pop. B only show a hard power/law. A comparison of all available (up to 2006)  XMM spectra ($n=$ 20 -- 40) for Pop. A and B sources \citep{sulenticetal08} confirms the differences originally reported in \citet{sulenticetal00b} in the sense that only Pop. A quasars show a soft excess. Note that the papers providing XMM data for this study consistently report no trends. SH14 show more Chandra/XMM data points than were available in 2006 but it is unclear what filters  have been applied to existing X-ray data (e.g number of hard/soft X-ray photons detected).

\subsection{RL/RQ dichotomy}

 There has been considerable work seeking connections between 4D Eigenvector 1
parameters and radio properties both to use  radio-loud sources as orientation  
indicators and to explore the RQ-RL dichotomy \citep{willsbrowne86,marzianietal01,rokakietal03,sulenticetal03,zamfiretal08}. \citetalias{shenho14} wisely distinguish between core and lobe dominant RL sources in their SFIX. 
The problem with this ``rediscovery" and a step backwards comes from their adopted 
definition of radio-loudness. A widely used criterion sees as RL all sources with 
radio/optical flux ratio \rk$>$10 \citep{kellermannetal89}. Our own work \citep{sulenticetal03,zamfiretal08} has focussed  on lobe-dominated (LD) RL sources as the parent population of classical RL quasars.  LD sources show a well defined lower limit in radio power (\rk$>$70) and a restricted occupation along the upper end of the 4DE1 main sequence. CD sources do not show these properties and those between \rk=10-70 distribute along the main sequence like  the RQ quasar majority. Inclusion of these sources in a RL sample will tend to  obscure differences (and dichotomy) between RQ and RL quasars. Perhaps ironically, a  larger sample of the brightest 150 LD quasars from SDSS DR9 confirm our original  lower radio power limit for these sources as well as  their restricted 4DE1  population B occupation.

\section{Important open issues}
\label{open}

 {The previous sections outline an undeniable progress in the understanding of quasar multifrequency properties, with a successful contextualisation in eigenvector 1 based schemes. We think that there are however some open issues that should be faced openly to set eigenvector 1 studies on a firmer ground. The first one is undeniably the formation of \feiiopt\ and \feiiuv\ lines. Current photoionization calculations do not allow to reproduce sources with \rfe $\gtrsim 1$, and it is still unclear whether this is a shortcoming of photoionization codes (that being based on a mean escape probability formalism for the treatment of radiation transfer, are not suited to consider low ionization line formation in the extended partially ionised zone where absorption processes are mainly non-local). On the one hand a study of other low ionization lines like the \caii\ IR triplet indicate that a dense, low ionization photoionizing medium can reproduce their total emission even in the most luminous sources \citep[][]{matsuokaetal07,martinez-aldamaetal15}. On the other hand, extreme Pop. A sources with \rfe $\gtrsim 1$\ are still poorly understood. They are most likely affected by a strong wind that may lead to shielding of the continuum \citep[e.g.,][]{elvis00,leighlymoore04,leighlyetal07}  or to mechanical heating  of the line emitting gas.  The relatively old paper by \citet{marzianietal01} estimated a decrease in $U$ with increasing \lledd, but connect $U$ and \feii\ prominence on an empirical relation and on crude scaling assumptions. Another aspect, that is severely constrained by data availability, is the interpretation of the \gs\ that may complicated by the presence of the warm absorber \citep{doneetal12,chakravortyetal12}. Last, we have to mention that not all workers agree that the \civ\ blueshift can be interpreted in terms of an outflow \citep{gaskellgoosmann08}, and that the best approach to extract a virial broadening indicator from the line width of \hb, especially in Pop. B,  is still a debated issue \citep[e.g.,][]{collinetal06,marzianisulentic12,shen13}. In this context, it remains important to ascertain the nature of the redward displacement of the VBC in Pop. B sources (Sulentic et al., in preparation). }
 
\section{Conclusion}

SH14  confirm  several empirical trends that were discovered 
earlier. Among them the ones involving \oiiiopt, distributional differences of RL/RQ 
sources in 4DE1, \hb\ profile properties as well as soft X-ray and CIV measures. SH14 adds 
an empirical verification that  orientation effects matter also for RQ sources -- something 
missing from earlier papers that  used the FWHM dependence on orientation derived from 
RL samples. It does not clarify the physical factors driving the location of sources in 
the 4DE1 optical plane. Only a multidimensional  approach involving at least \mbh, \lledd\ 
and orientation \citep[cf.][]{laor00,marzianietal01,zamanovetal02} can lead to a final clarification of these issues and successful 
development of a physical model. 

\pagebreak
\bigskip

\begin{figure}[htp!]
\centering
\includegraphics[width=3.1in]{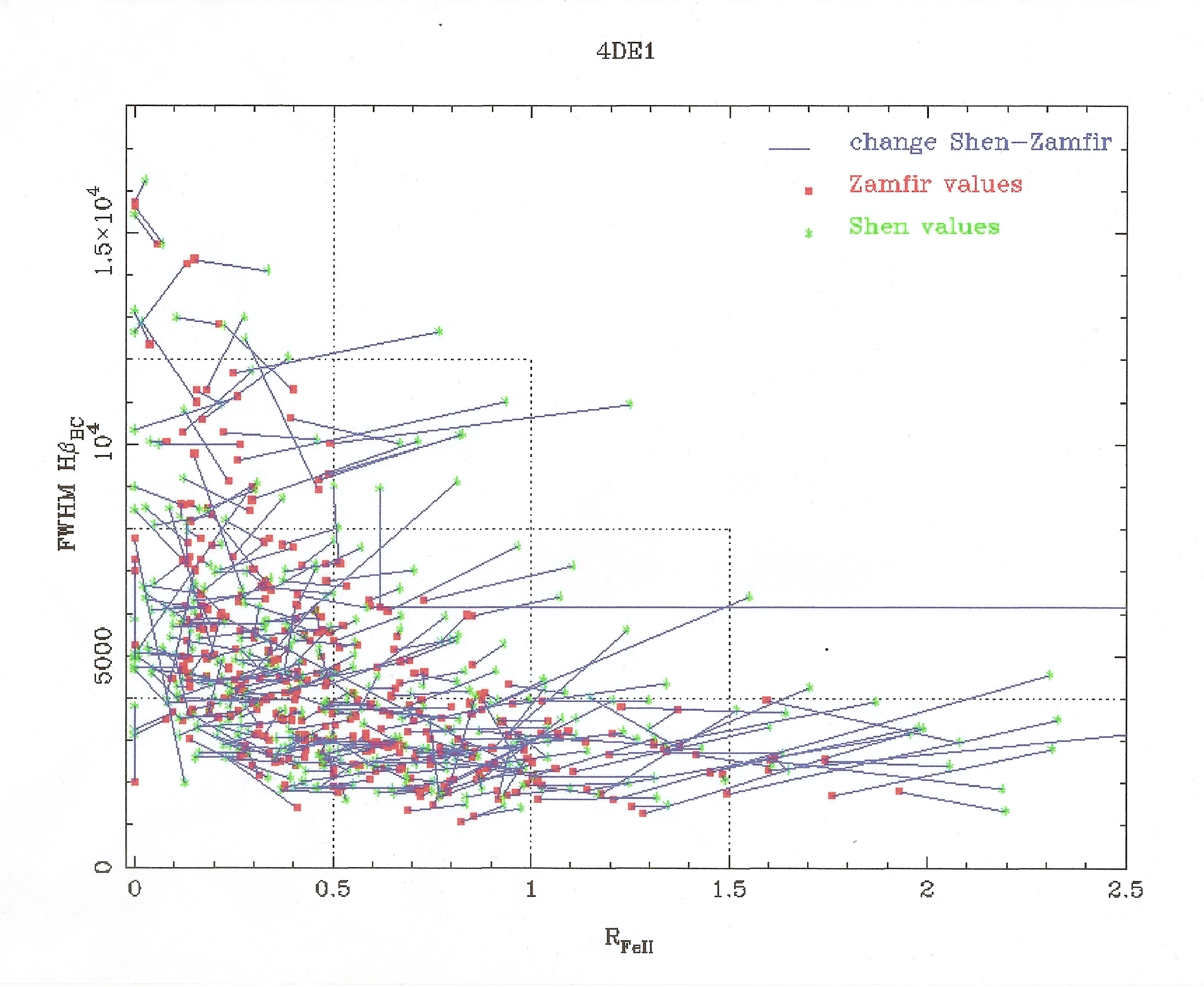}
\includegraphics[width=3.1in]{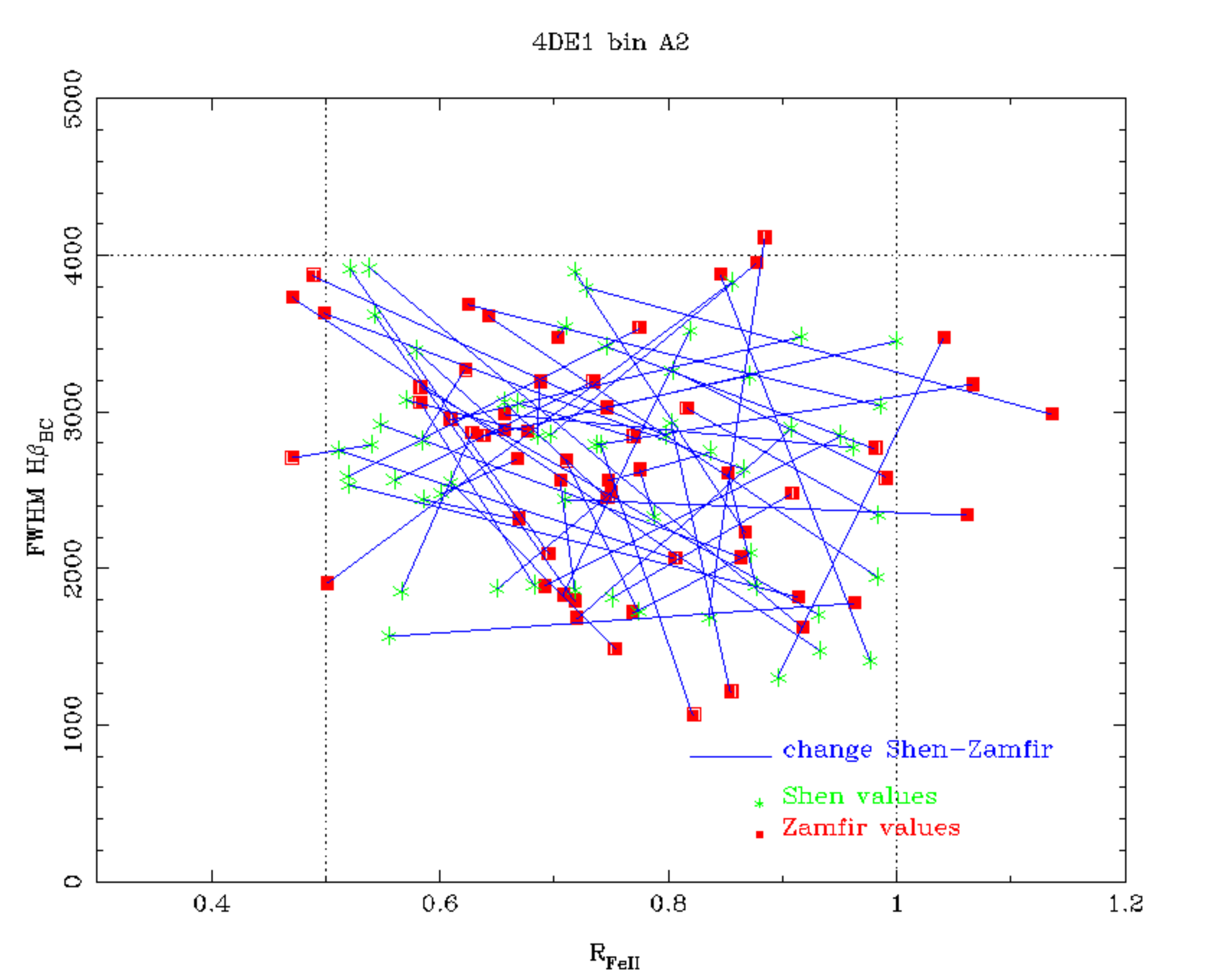}
\caption{Left: a comparison between \citet{shenetal11} and \citet{zamfiretal10} in the optical plane of 4DE1  involving the width of H$\beta$ and the strength of optical \feii. The blue lines connect \citet{zamfiretal10} (green stars) with \citet{shenetal11} measurements for the same objects. Right: A2 and part of adjacent bins in the 4DE1 optical plane \label{fig:shz}} 
\end{figure}

\begin{figure}[htp!]
\centering
\includegraphics[width=3.1in]{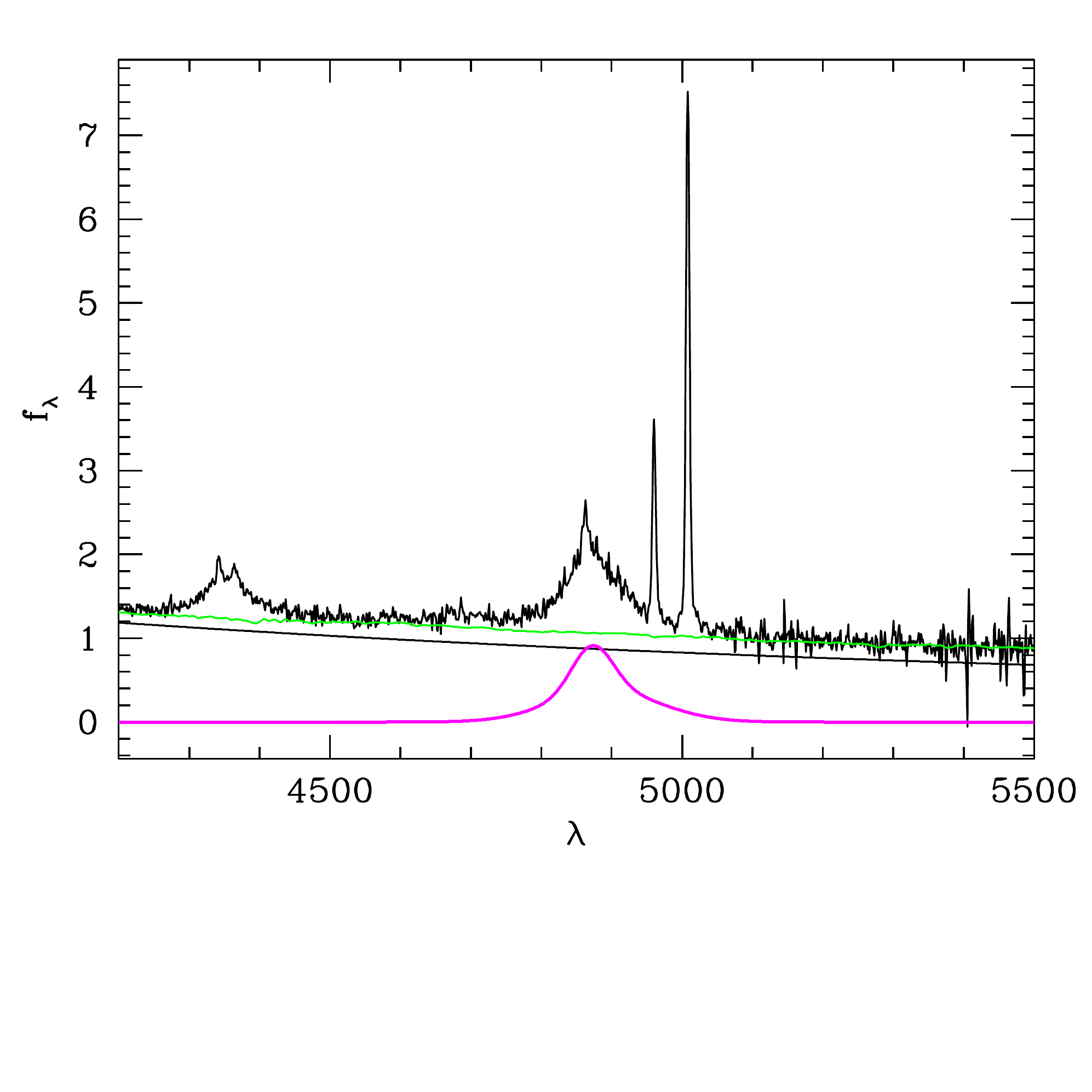}\\
\includegraphics[width=3.1in]{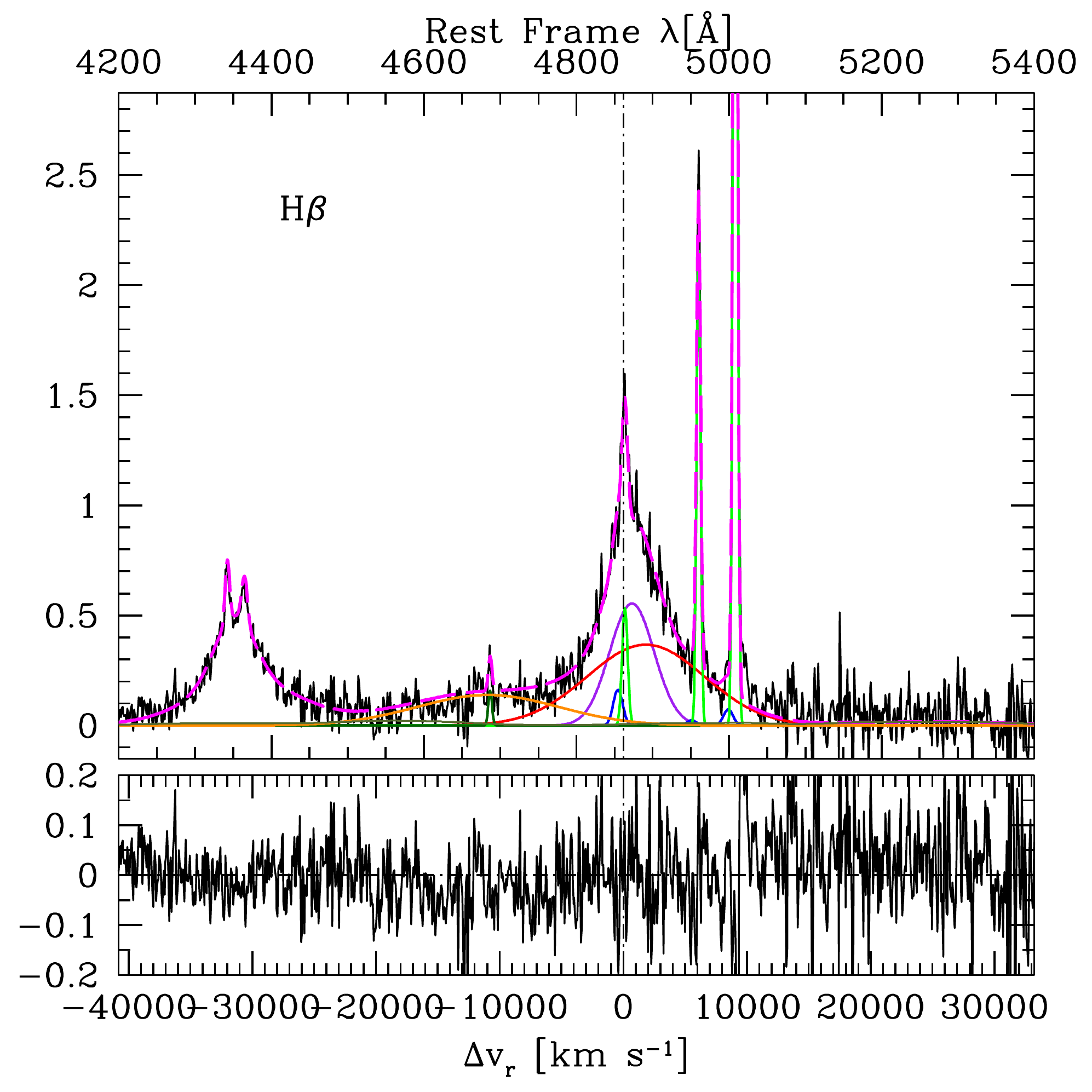}
\caption{ {Top: continuum placement  for source SDSS J034106.75+004609.9.  The black line traces the featureless continuum; the green line traces the total continuum that includes a contribution from the host galaxy (whose presence is made manifest by the Mg Ib band visible in the spectrum. There is no significant \feii\ emission.  The magenta line traces the cleaned  \hb\ broad profile. Bottom: \hb\ profile analysis. The   excess on the blue side of \hb\ can be explained by significant \heii\ VBC emission. \hb\ and \hg\ are consistently fit with the same BC and VBC parameters (the BC and VBC (red) decomposition is shown for \hb\ only). }  \label{fig:heii}} 
\end{figure}

\begin{figure}[htp!]
\centering
\includegraphics[width=6.1in]{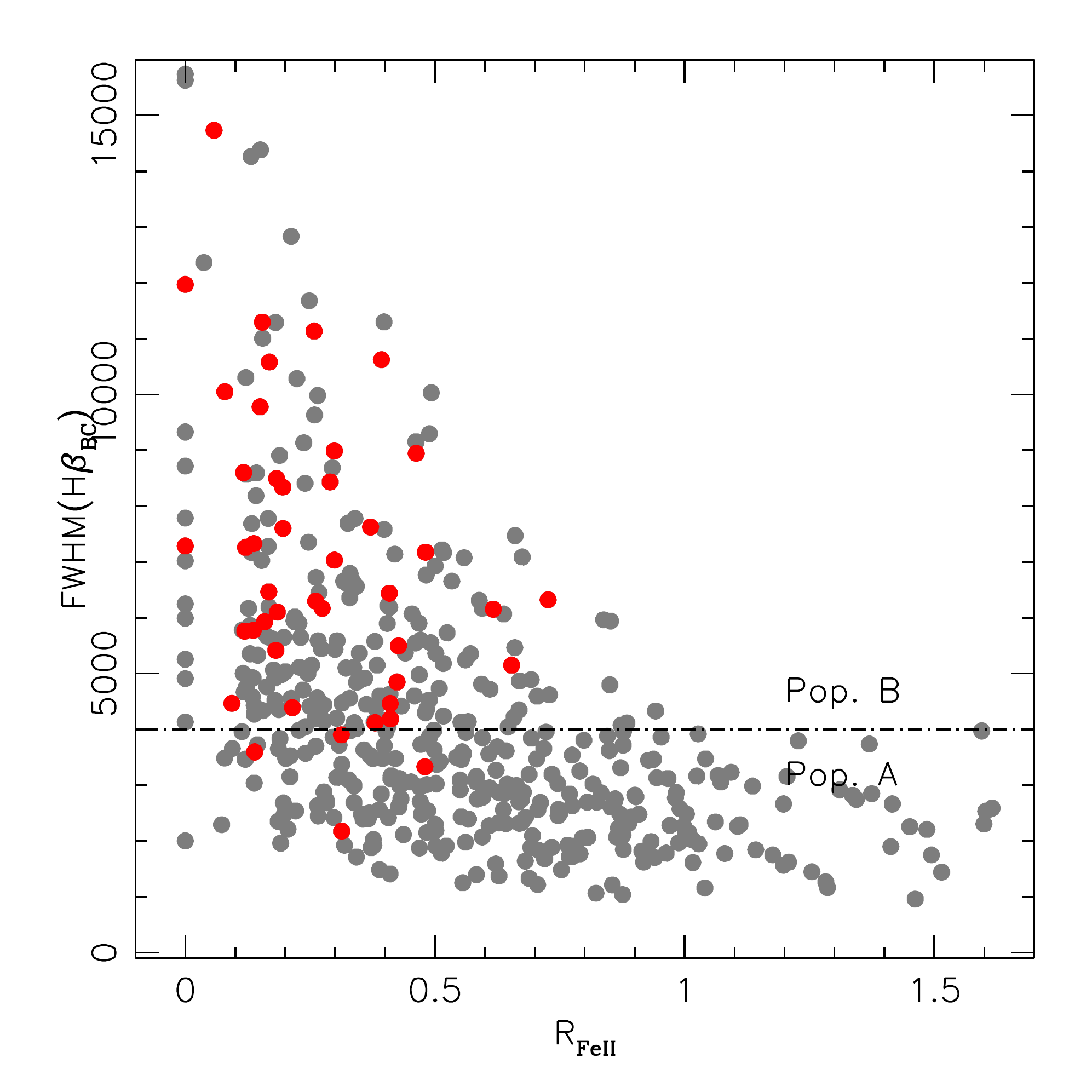}
\caption{The 4DE1 optical plane involving the width of H$\beta$ and the strength of optical FeII. 
Horizontal line  marks the boundary between population A (lower) and B (upper) sources. Source 
occupation is shown for the 470 brightest SDSS-DR5 quasars (z$<$0.75) with highest s/n SDSS 
spectra \cite{zamfiretal08}. Grey and red dots represent RQ and lobe dominated LD RL quasars.
The latter show a strong preference for the Population B zone.\label{fig:z08} } 
\end{figure}

\begin{figure}[htp!]
\centering
\includegraphics[width=6.1in]{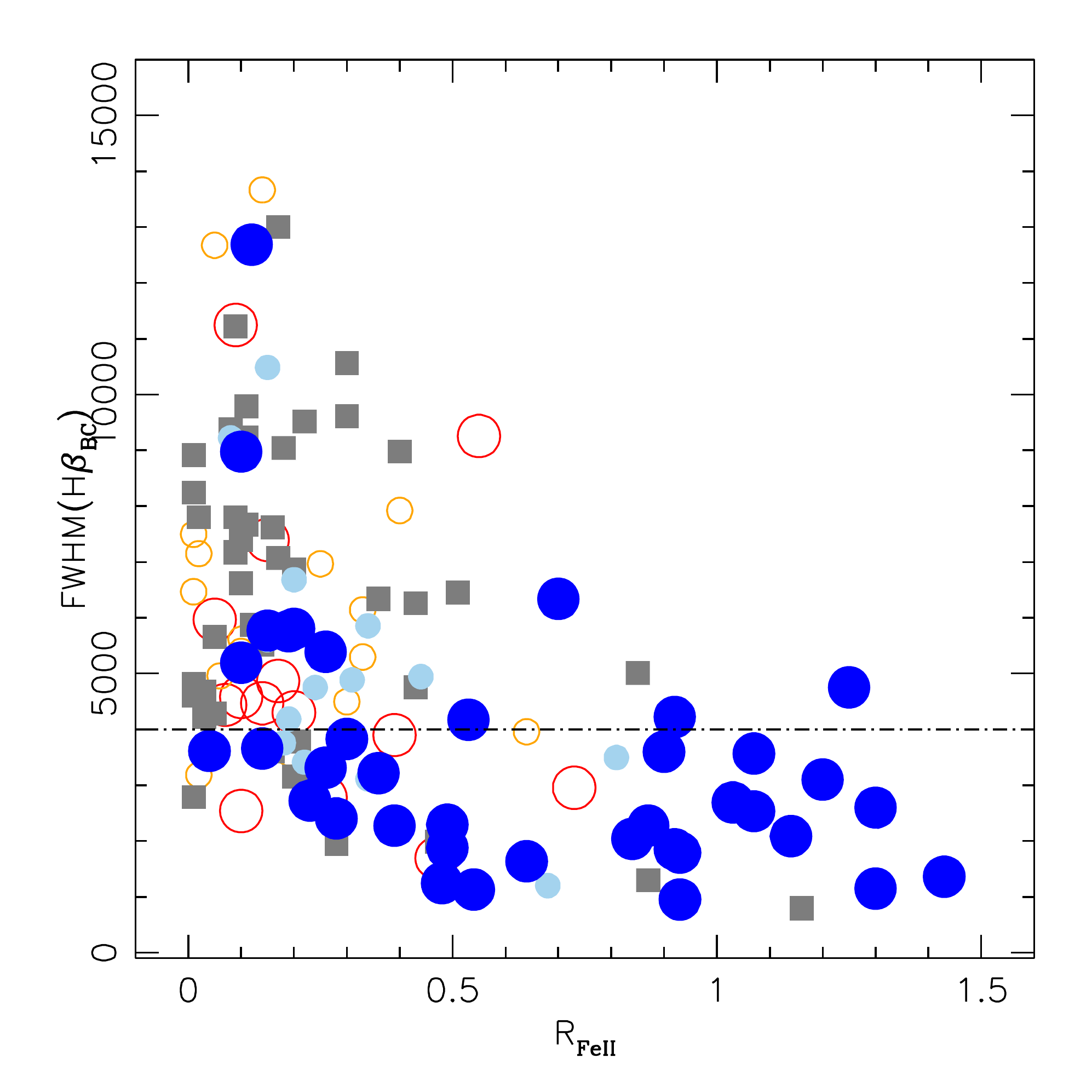}
\caption{Same optical plane as the previous Figure, for all low z quasars with measurable 
HST/FOS UV spectra \citep{sulenticetal07}. Different symbols represent the amplitude of 
the CIV 1549 blueshift at half maximum $c(\frac{1}{2})$. Large blue filled circles
involve sources with largest CIV blueshifts which strongly favor the Population A zone.  
Large open red circles represent sources with largest \civ\ redshift and grey squares those 
with no significant line shift \citep{sulenticetal07}. \label{fig:s07}} 
\end{figure}

\begin{figure}[htp!]
\centering
\includegraphics[width=4.1in, angle=-0]{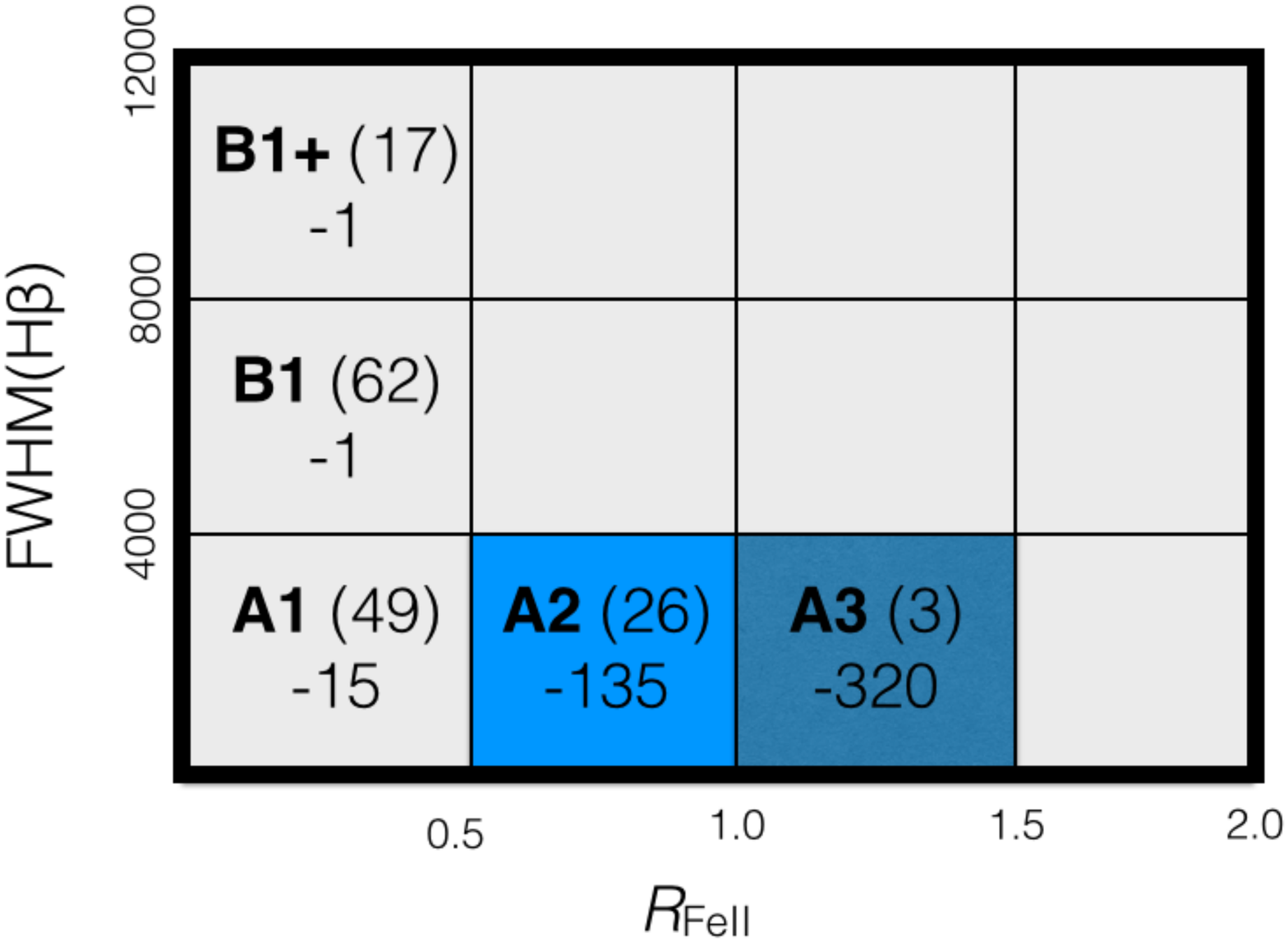}
\caption{Average shift of \oiii\ (negative is blueshift) for spectral types in the optical plane of E1. 
Numbers in parentheses indicate sources in each bin, with the dataset available to us in late 2001. \label{fig:bo}} 
\end{figure}

I special thank you to Chony del Olmo for the production of Fig. \ref{fig:shz}. We wish also to thank the many collaborators, post-docs and former students that, over almost two decades, have contributed to the development of the eigenvector 1 ideas: Deborah Dultzin,  Tomaz Zwitter, Giovanna M. Stirpe, Rumen Bachev,   Radoslav Zamanov,  Chony del Olmo, Mauro D'Onofrio, Alenka Negrete, Massimo Calvani, Patricia Romano, Mary Loli Martinez Aldama, Ilse Plauchu-Frayn. 



\vfill\eject\pagebreak

\newpage\vfill\eject\pagebreak

\begin{multicols}{2}\small

\expandafter\ifx\csname natexlab\endcsname\relax\def\natexlab#1{#1}\fi
\end{multicols}

\end{document}